# SUGAR: Spherical Ultrafast Graph Attention Framework for Cortical Surface Registration

Jianxun Ren[1,*], Ning An[1,*], Youjia Zhang[1], Danyang Wang[1], Zhenyu Sun[1], Cong Lin[1], Weigang Cui[2], Weiwei Wang[1], Ying Zhou[1], Wei Zhang[1,3], Qingyu Hu[1], Ping Zhang[1], Dan Hu[4], Danhong Wang[4], Hesheng Liu[1,5,#]

[1] Changping Laboratory, Beijing, China

[2] School of Engineering Medicine, Beihang University, Beijing, China

[3] Academy for Advanced Interdisciplinary Studies, Peking University, Beijing, China

[4] Athinoula A. Martinos Center for Biomedical Imaging, Department of Radiology, Massachusetts General Hospital, Harvard Medical School, Charlestown, MA, USA

[5] Biomedical Pioneering Innovation Center (BIOPIC), Peking University, Beijing, China

* These authors equally contributed to this work

# Address correspondence to:
Dr. Hesheng Liu,
Changping Laboratory, Beijing, China
Email: liuhesheng@cpl.ac.cn

**Abstract**

Cortical surface registration plays a crucial role in aligning cortical functional and anatomical features across individuals. However, conventional registration algorithms are computationally inefficient. Recently, learning-based registration algorithms have emerged as a promising solution, significantly improving processing efficiency. Nonetheless, there remains a gap in the development of a learning-based method that exceeds the state-of-the-art conventional methods simultaneously in computational efficiency, registration accuracy, and distortion control, despite the theoretically greater representational capabilities of deep learning approaches. To address the challenge, we present SUGAR, a unified unsupervised deep-learning framework for both rigid and non-rigid registration. SUGAR incorporates a U-Net-based spherical graph attention network and leverages the Euler angle representation for deformation. In addition to the similarity loss, we introduce fold and multiple distortion losses, to preserve topology and minimize various types of distortions. Furthermore, we propose a data augmentation strategy specifically tailored for spherical surface registration, enhancing the registration performance. Through extensive evaluation involving over 10,000 scans from 7 diverse datasets, we showed that our framework exhibits comparable or superior registration performance in accuracy, distortion, and test-retest reliability compared to conventional and learning-based methods. Additionally, SUGAR achieves remarkable sub-second processing times, offering a notable speed-up of approximately 12,000 times in registering 9,000 subjects from the UK Biobank dataset in just 32 minutes. This combination of high registration performance and accelerated processing time may greatly benefit large-scale neuroimaging studies.

**Keywords:** cortical surface registration, graph neural network, attention mechanism, distortion, big data

# 1. Introduction

Surface-based analyses have become increasingly popular in anatomical and functional neuroimaging research (Coalson et al., 2018; Fischl, 2012; Van Essen, 2005). The surface mesh offers a suitable representation for the complex morphological features of the cerebral mantle (Dale et al., 1999; Van Essen et al., 2001; Van Essen et al., 1998). Such analyses thus enable more accurate cross-individual alignment of functional and anatomical features in the cerebral cortex through cortical surface registration, compared to volumetric approaches (Anticevic et al., 2008; Coalson et al., 2018; Ghosh et al., 2010; Jo et al., 2008; Oosterhof et al., 2011; Tucholka et al., 2012). Typically, cortical surface registration involves two steps: rigid and non-rigid registration. Rigid registration employs a global linear transformation to achieve coarse morphological alignments, typically taking minutes to process. Subsequently, non-rigid registration performs vertex-wise local deformations to achieve fine-grained and more accurate alignments. However, the high computational complexity associated with non-rigid registration results in processing time of dozens of minutes, such as over 30 minutes in the case of FreeSurfer (Fischl et al., 1999a; Fischl et al., 1999b). This computational inefficiency poses challenges for downstream analyses, such as anatomical parcellation and group-average analyses, particularly when handling large-scale datasets (Fischl et al., 2004; Yeo et al., 2011). Furthermore, the non-rigid deformations, solely aimed at improving the registration accuracy, may lead to topological errors and implausible local distortions (Robinson et al., 2014; Yeo et al., 2009). Hence, striking a balance among increasing registration accuracy, preserving topology, and minimizing registration distortion is of paramount importance, albeit challenging. Consequently, the evaluation of registration can be summarized based on four key aspects: computational efficiency, registration accuracy, topology preserving, and registration distortion.

FreeSurfer has been at the forefront of introducing widely-used rigid and non-rigid cortical surface registration methods for the past two decades (Fischl et al., 1999a; Fischl et al., 1999b). However, the computational efficiency of non-rigid registration in FreeSurfer is limited, as aforementioned. Spherical Demons (SD) has made significant strides in enhancing efficiency for the non-rigid registration, completing the registration process in less than 5 minutes while preserving topology and ensuring high levels of accuracy and distortion control (Yeo et al., 2009).

Multimodal Surface Matching (MSM) achieved surface registration by leveraging multimodal features beyond a single geometric aspect, with one of its variants demonstrating state-of-the-art distortion control (Robinson et al., 2018; Robinson et al., 2014). Nevertheless, MSM comes with a high computational burden.

Recently, registration algorithms based on deep learning have greatly improved the computational efficiency of non-rigid volumetric and surface registration while delivering comparable performance to the conventional approaches (Balakrishnan et al., 2019; Cheng et al., 2020; Dalca et al., 2019; Suliman et al., 2022a; Suliman et al., 2022b; Zhao et al., 2021a). For the cortical surface registration, SphereMorph parameterizes spherical meshes into a 2D rectangular image through planar projection (Cheng et al., 2020). Subsequently, a convolutional neural network (CNN) is employed to achieve comparable accuracy to FreeSurfer and SD with a processing time of less than 1 minute. However, the parameterization used in SphereMorph introduces topological errors and distortions. In contrast, S3Reg employs hexagonal filters of the Spherical U-Net, a spherical CNN architecture (Zhao et al., 2019), to achieve direct spherical convolution, thus reducing distortions induced by parametrization (Zhao et al., 2021a). In addition, S3Reg preserves topology and demonstrates a further improvement in processing time of less than 10 seconds. However, due to the lack of a global coordinate system, the local coordinates of hexagonal filters flip across spherical poles causing polar distortions. To address this limitation, S3Reg attempts to incorporate three orthogonal spheres. On the other hand, Deep-Discrete Learning Framework (DDR) and GeoMorph utilize MoNet, a generic graph neural network (GNN) applicable to non-Euclidean structured data (Monti et al., 2017), to address polar distortions (Suliman et al., 2022a; Suliman et al., 2022b). Unlike the continuous registration framework of SphereMorph and S3Reg, DDR and GeoMorph adapt a discrete registration framework following the ideas of MSM. By incorporating smoothing and as-rigid-as-possible regularizations, GeoMorph generates smooth surfaces with minimal distortions, and it avoids self-intersections by limiting its deformation stride, thus yields fewer distortions than S3Reg but still more than SD, while maintaining the same level of registration accuracy. However, it falls short in simultaneously achieving higher registration accuracy, lower distortion, and shorter processing time compared to the state-of-the-art conventional approach, leaving a gap in the development of a learning-based registration method that can sufficiently harness the potential of deep learning frameworks.

Here, we propose a continuous spherical ultrafast graph attention framework for cortical surface registration (SUGAR). The framework incorporates a U-Net-based spherical graph attention network (S-GAT) as the architecture. The S-GAT is built upon the graph attention network (GAT) that incorporates attention mechanisms within the GNN (Brody et al., 2021; Velickovic et al., 2017; Zhou et al., 2020). These attention mechanisms assign varying weights to neighboring vertices, enabling the network to demonstrate better performance than the MoNet (Velickovic et al., 2017). Based on the S-GAT, we utilize Euler angles to represent deformations and achieve both rigid and non-rigid registration within a unified framework. Moreover, in addition to the similarity term in the loss function, we also explicitly introduce a fold loss to preserve topology and three distortion losses to constrain various types of distortions: shape, area, and edge distortions. The design of loss functions facilitates the flexible trade-off between registration accuracy and distortion. Furthermore, while previous learning-based methods train models by aligning subject spheres to a fixed atlas, we propose a data augmentation strategy for the cortical surface registration task. This strategy randomly selects and rotates subject's spheres from the training set as target spheres, benefiting the model to learn complex idiosyncratic features and thus enhancing registration performance. We perform extensive registration experiments using our proposed method on over 10,000 scans from seven different datasets, achieving more satisfactory registration performance in terms of computational efficiency, accuracy, and distortion compared to both state-of-the-art conventional methods and the existing learning-based approach.

The main contributions of the paper are summarized below:
1. We propose a unified continuous framework, SUGAR, for both rigid and non-rigid cortical surface registration, which integrates the attention mechanism into a spherical graph neural network, improving the overall registration performance.
2. We introduce fold and distortion losses in the loss function to guarantee topology preserving and minimize distortions of deformations, balancing the registration accuracy and distortions.
3. We propose a data augmentation strategy for spherical surface registration, involving random replacement and rotation of spherical surfaces, leading to improved and robust subject-to-atlas and subject-to-subject registration performance.
4. We extensively evaluate the performance of SUGAR in over 10,000 scans from seven diverse

datasets and demonstrate comparable if not better performance in accuracy, distortions, and test-retest reliability, against other conventional and learning-based methods, as well as accelerating processing time to the sub-second level.

## 2. Method

In cortical surface registration, mapping cortical surfaces into spherical meshes is a commonly used method, which naturally preserve the topology. The surface registration process involves deforming a moving sphere ($M$) to align with a fixed sphere ($F$), while simultaneously maximizing similarities, preserving topology, and minimizing distortions. Both $M$ and $F$ consist of $N$ vertices $\{v_i\}_{i=1}^{N}$, where $v_i \in \mathbb{R}^3$ is represented in the Cartesian coordinate (i.e. $xyz$-coordinate).

### 2.1. Overall Framework

SUGAR is designed for both rigid and non-rigid registration in sequence, as illustrated in Figure 1. Specifically, a mapping function, denoted as $f: \mathbb{R}^3 \to S^2$, is utilized to transform vertices from Cartesian to spherical coordinates, where a vertex on the sphere is represented as $(\rho, \theta) = f(x, y, z)$. These vertices are subsequently encoded using a positional encoding (PE) technique (detailed in Section 2.3) to map them to a higher-dimensional feature space. The encoded positional features $(\eta(\rho), \eta(\theta))$, along with $(\rho, \theta)$ and geometric features such as sulcal depth, are then fed into the S-GAT (denoted as $g_\xi(M, F)$) to generate a set of global Euler angles ($\varphi_{rigid} \in \mathbb{R}^{1 \times 3}$) through a global averaging pooling layer for rigid registration, or vertex-wise Euler angles ($\varphi_{non-rigid} \in \mathbb{R}^{N \times 3}$) for non-rigid registration. Here, $\xi$ represents the learnable parameters and $N$ denotes the number of cortical vertices.

In the subsequent Spatial Transform Layer (STL), these Euler angles are transformed into rotation tensors ($\Phi_{rigid} = R(\varphi_{rigid}) \in \mathbb{R}^{1 \times 3 \times 3}$ for rigid registration, and $\Phi_{non-rigid} = R(\varphi_{non-rigid}) \in \mathbb{R}^{N \times 3 \times 3}$ for non-rigid registration). These rotation tensors are then applied to warp the moving sphere $M$ into a moved sphere $M \circ \Phi$. For rigid registration, a similarity term is employed to maximize alignment between input pairs. The moved sphere after the rigid registration replaces the original $M$ as the input for the non-rigid registration. In non-rigid registration, an additional regularization term is introduced to preserve topology and minimize distortion while ensuring high alignment accuracy. To improve anatomical parcellation accuracy, a parcellation term utilizing anatomical parcellation is incorporated (Balakrishnan et al., 2019; Zhao et al., 2021a). Moreover, a multi-resolution approach is employed during training, using four different resolutions of subdivided icosahedrons (icospheres) to reduce computational burden, including $ico_3$, $ico_4$, $ico_5$, and $ico_6$ corresponding to the number of vertices of 642, 2562, 10242, and 40962, respectively (Yeo et al., 2009; Zhao et al., 2021a).

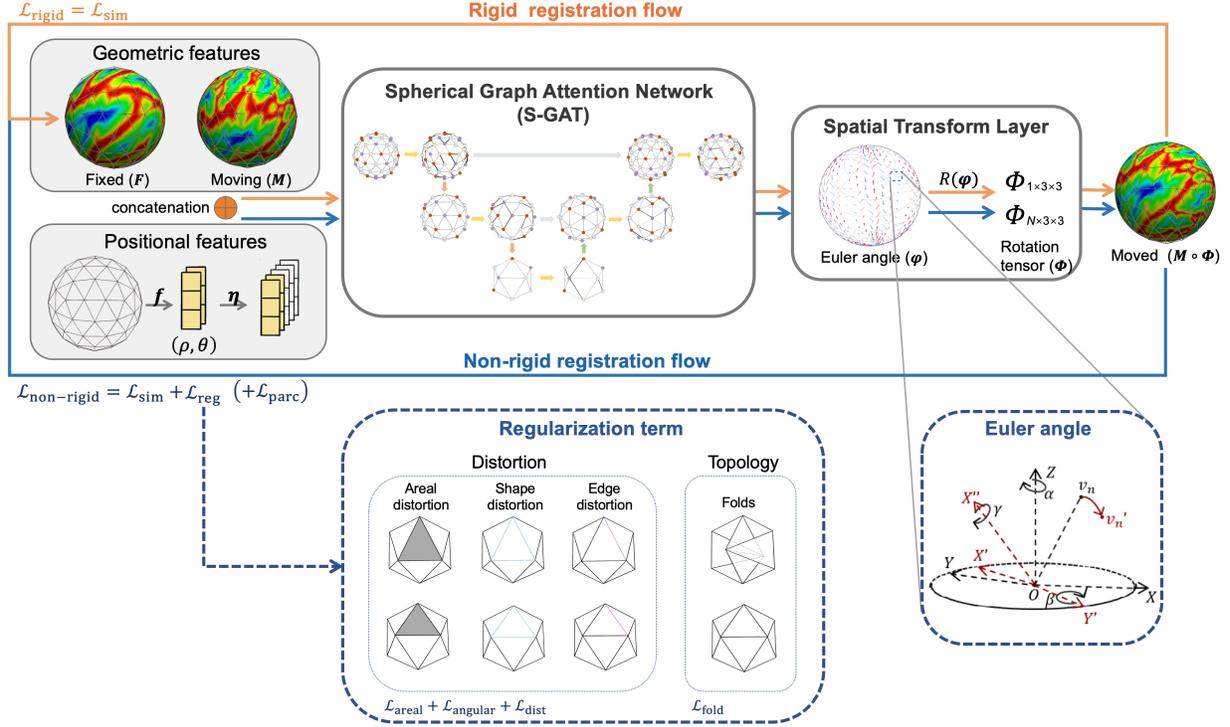

**Figure 1. Illustration of the SUGAR framework.** The model inputs include the geometric features and Cartesian coordinates of the Moving ($M$) and Fixed ($F$) icospheres. The Cartesian coordinates are transformed ($f$) into spherical coordinates and further encoded ($\eta$) as higher-dimensional positional features. These geometric features and positional features are fed into the spherical graph attention network (S-GAT). The S-GAT outputs a set of global Euler angle $\varphi$, which are transformed into a rotation tensor $\Phi = R(\varphi)$ in the spatial transform layer to rigidly align the input pairs, indicated by the orange flow. Subsequently, the rigidly registered Moved ($M \circ \Phi$) sphere replaces the original $M$ as the input for the subsequent non-rigid registration, indicated by the blue flow. In the rigid registration, only similarity term was used in the loss function. In the non-rigid registration, three types of distortion losses and a fold loss are applied in additional to the similarity term to constrain various distortions and eliminates folds.

## 2.2 S-GAT

The spatial approaches of GNNs define operations directly on nodes and their neighbors on the graph domain, making them inherently compatible with the data structure of surface meshes. One popular variant of GNNs is the GAT that incorporates attention mechanisms (Brody et al., 2021; Vaswani et al., 2017; Velickovic et al., 2017; Yun et al., 2019; Zhang et al., 2020). Unlike other GNN variants weight all edges equivalently, GATs assign varying weights to each edge for every node by attending to its neighborhood. In this work, we extend the capabilities of graph networks to cortical spherical meshes and propose the S-GAT that adopts a U-Net architecture (Figure 2). The S-GAT consists of multiple layers that enable effective information propagation and integration across the spherical mesh, including the graph attentional layer, activation layer, pooling layer, unpooling layer and skip-connection strategy.

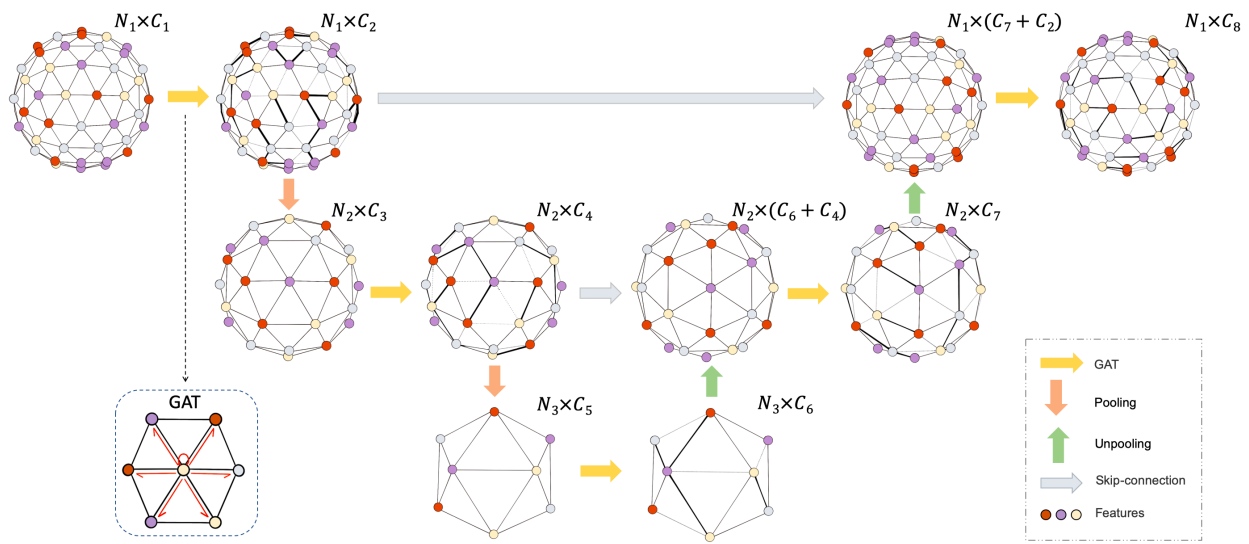

**Figure 2. S-GAT architecture.** The S-GAT is structured as a U-Net architecture. The GAT layer (yellow arrows) within the S-GAT learns the weights of the input features by attending to every node and edge in the icosphere. The current icosphere is then downsampled to the lower-resolution icosphere through the pooling layer (orange arrows). Once the S-GAT captures the most salient features, the icosphere begins to be upsampled to the higher-resolution icosphere through the unpooling layer (green arrows). These upsampled features are concatenated with the features from the corresponding pooling layer using a skip-connection strategy (gray arrows), iteratively restoring the original resolution of the icosphere. Finally, the S-GAT outputs a set of Euler angles. Each dot with different colors represents each node with different features. $N_i$ denotes the number of vertices in the icosphere $ico_i$ and $C_j$ denotes the number of channels.

***Graph attentional layer.*** The inputs of a single graph attentional layer are a set of node features $H = \{h_1, h_2, \ldots, h_N\}, h_i \in \mathbb{R}^F$ and the output is $H' = \{h'_1, h'_2, \ldots, h'_N\}, h'_i \in \mathbb{R}^{F'}$, where $N$, $F$, and $F'$ denote the number of vertices and feature dimensions for inputs and outputs.

To determine the importance of features of node $j$ to node $i$, we compute an attention coefficient using an approach proposed by (Brody et al., 2021), which employs self-attention mechanism of the Transformer (Vaswani et al., 2017) and applies a weight matrix $W \in \mathbb{R}^{F' \times 2F}$ to every node:

$$e(h_i, h_j) = \mathbf{a}^T \text{LeakyReLU}(W[h_i \parallel h_j]), \tag{1}$$

where $j \in \mathcal{N}_i \cup \{i\}$ represents the closest nodes to a given node $i$, forming a 1-hop neighborhood ($\mathcal{N}_i$) and including itself $i$. Here, $\parallel$ denotes the concatenation operation, and $\mathbf{a} \in \mathbb{R}^{F'}$ denotes a weight vector of a linear transformation. Since attention coefficients are independently calculated for each pair of neighboring nodes, they are normalized across the neighborhood using a softmax function for direct comparison,

$$\alpha_{ij} = \text{softmax}_{j \in \mathcal{N}_i}\left(e(h_i, h_j)\right) = \frac{\exp\left(e(h_i, h_j)\right)}{\sum_{k \in \mathcal{N}_i \cup \{i\}} \exp(e(h_i, h_k))}. \tag{2}$$

We used the multi-head attention mechanism (Vaswani et al., 2017) that incorporates $K$ independent attention heads in parallel, enabling a more stable training process and further improving accuracy. The final transformed weighted output features for each node is

$$h'_i = \parallel_{k=1}^{K} \sum_{j \in \mathcal{N}_i \cup \{i\}} \alpha_{ij}^k W^k h_j. \tag{3}$$

***Activation layer.*** The output feature $h'_i$ from the graph attentional layer is subjected to a nonlinear activation function $\sigma$ in the activation layer, resulting in the transformed feature $h'_i = \sigma(h'_i)$. In this case, the LeakyReLU function is used as the nonlinear activation function.

***Pooling layer.*** The pooling operation is performed on an icosphere at a resolution of $\text{ico}_{l-1}$. The feature $H_{l-1}$ of each vertex from the icosphere $\text{ico}_{l-1}$ are updated by averaging the features across itself and its 1-hop neighborhood $\mathcal{N}_i \cup \{i\}$ at the icosphere $\text{ico}_l$. To accomplish this, we used the scatter mean, a sparse operation commonly used in graph-based methods, rather than the matrix mean from the spherical surface pooling, which requires predefined neighborhood orders

(Zhao et al., 2019). The scatter mean operation is independent on the predefined orders, making it more adaptable to non-standard spheres, such as individual-specific cortical surfaces.

***Unpooling layer and skip-connection strategy.*** In the unpooling operation, the feature $H_{l+1}$ of each vertex from the icosphere $ico_{l+1}$ are updated by averaging the features $H_l$ across the two neighboring vertices at the icosphere $ico_l$. Additionally, the resulting feature representations of $H_{l+1}$ are concatenated with the features from the corresponding pooling layer using a skip-connection strategy (Ronneberger et al., 2015).

The final graph attentional layer yields a set of Euler angles $\varphi \in \mathbb{R}^{N \times 3}$, where $N$ represents the number of vertices for non-rigid registration and $N = 1$ for rigid registration.

### 2.3 Positional encoding

The three-dimensional Cartesian coordinate $(x, y, z)$ of a vertex on a unit sphere can be parameterized using two angles: an azimuthal angle $\rho = \mathrm{atan2}\left(\frac{y}{x}\right) \in [-\pi, \pi]$ and a polar angle $\theta = \mathrm{atan2}\left(\frac{\sqrt{x^2+y^2}}{z}\right) \in [0, \pi]$. They are normalized to $\rho = \frac{\rho}{2\pi} + \frac{1}{2} \in [0,1], \theta = \frac{\theta}{\pi} \in [0,1]$. The spherical coordinates $(\rho, \theta)$ of each node is encoded using the PE strategy to augment positional features (Gao, 2023; Mildenhall et al., 2021), which is based on a set of sinusoidal functions that maps a single spherical coordinate $\mathrm{pos} \in \mathbb{R}$ to a high-dimensional space $\eta(\mathrm{pos}) \in \mathbb{R}^{2L+1}$,

$$\eta(\mathrm{pos}) = (\sin(2^0\pi \cdot \mathrm{pos}), \cos(2^0\pi \cdot \mathrm{pos}), \ldots, \sin(2^{L-1}\pi \cdot \mathrm{pos}), \cos(2^{L-1}\pi \cdot \mathrm{pos}), pos). \quad (4)$$

The feature mapping $\eta(\cdot)$ is separately applied to both normalized spherical coordinates $\rho$ and $\theta$, and $L$ is empirically set to 4. The encoded positional features of each vertex are unique and contain relative positional information. The encoded positional features along with geometric features are fed into the S-GAT.

### 2.4 Spatial transform layer

The STL transforms output Euler angles ($\varphi$) from S-GAT into a rotation tensor ($\Phi$). Specifically, a single rotation matrix $\Phi_i$ of the vertex $i$ from the rotation tensor is defined as:

$$\begin{aligned}\Phi_i = R(\varphi_i) &= Z(\gamma)Y'(\beta)X''(\alpha) \\ &= \begin{bmatrix} \cos\alpha\cos\beta & \cos\alpha\sin\beta\sin\gamma - \cos\gamma\sin\alpha & \sin\alpha\sin\gamma + \cos\alpha\cos\gamma\sin\beta \\ \cos\beta\sin\alpha & \cos\alpha\cos\gamma + \sin\alpha\sin\beta\sin\gamma & \cos\gamma\sin\alpha\sin\beta - \cos\alpha\sin\gamma \\ -\sin\beta & \cos\beta\sin\gamma & \cos\beta\cos\gamma \end{bmatrix}, \end{aligned} \quad (5)$$

where $\alpha, \beta, \gamma$ are Euler angles of three elemental rotations w.r.t. $Z - Y' - X''$ axis. By applying $\Phi$, we performed the spatial transformation on $M$ to obtain the moved sphere $M \circ \Phi$.

### 2.5 Optimization

Our loss function compromised a similarity term, a regularization term, and a parcellation term. During the optimization process, the feature similarity between $M \circ \Phi$ and $F$ is estimated using the Mean Squared Error (MSE) for both rigid and non-rigid registrations:

$$\mathcal{L}_{\text{sim}} = MSE(M \circ \Phi, F) = \frac{1}{N}\sum(M \circ \Phi - F)^2, \tag{6}$$

where $N$ is the number of vertices. However, simply improving similarities may yield implausible deformations without preserving topology and constraining mesh distortions. Here, we adapt the areal, angular and distance losses to minimize areal, shape and edge distortions and a fold loss to preserve mesh topology.

The areal loss is adapted from the one implemented in FreeSurfer (Fischl et al., 1999b). Instead of directly constraining the change in area for each triangle relative to its original state, we adopt a normalized version that constrains the ratio of area change, taking into account the variations in triangular areas at different resolutions and among individual spheres:

$$\mathcal{L}_{\text{areal}} = \frac{1}{T}\sum_{t=1}^{T}\left|1 - \frac{A_t}{A_t^0}\right|, \tag{7}$$

where $T$ denotes the number of triangles, $A_t^0$ and $A_t$ denote the original area and the area after deforming the triangle $t$, respectively.

Similarly, inspired by the areal loss, we propose an angular loss:

$$\mathcal{L}_{\text{angle}} = \frac{1}{3T}\sum_{t=1}^{T}\sum_{\tau=1}^{3}\left|1 - \frac{\angle t_\tau}{\angle t_\tau^0}\right|, \tag{8}$$

where $\angle t_\tau$ and $\angle t_\tau^0$ represent the current and the original $\tau^{\text{th}}$ angle of the triangle $t$, respectively.

To thoroughly restrict the overall distortion, we further penalize the relative distance change between centroids and barycenters of polygons. Most nodes are connected to six neighbors forming hexagons, except for twelve nodes that are connected to five neighbors, forming pentagons. These polygons approximate uniform hexagons and uniform pentagons and hence theoretically have their centroids coincide with the barycenters, but this is not the case in practice.

To address this bias, we design the following loss to maintain a consistent distance between the centroids and barycenters before and after deformations:

$$\mathcal{L}_{\text{dist}} = \frac{|d(\boldsymbol{v}'_C - \boldsymbol{v}'_B) - d(\boldsymbol{v}^0_C - \boldsymbol{v}^0_B)|}{\epsilon_l}, \quad (9)$$

where $d(\cdot)$ denotes the Euclidean distance, and $\epsilon_l$ is the mean edge length at the current $ico_l$ level. Moreover, $\boldsymbol{v}^0_C$ and $\boldsymbol{v}^0_B$ denote the original coordinates of the centroid and the barycenter of a 1-hop neighborhood, respectively. And $\boldsymbol{v}'_C$ and $\boldsymbol{v}'_B$ denote the updated coordinates after deformations.

Since the distortion constraints cannot completely prevent from self-intersections or folds of triangular faces, we further introduce the fold loss proposed by FreeSurfer (Fischl et al., 1999a). To distinguish whether a triangular face is folded or not, we constructed an oriented area of triangle $t$, $\Delta_t = \frac{\boldsymbol{v}_{i,j} \times \boldsymbol{v}_{i,k} \cdot \mathbf{n}}{2}$, where $\boldsymbol{v}_{i,j}$ and $\boldsymbol{v}_{i,k}$ denote the edge vectors from the node $i$ to other node $j$ and node $k$ of the triangle $t$, respectively. The unit normal vector $\mathbf{n}$ of the triangle points outward from the origin of the sphere. If $\boldsymbol{v}_{i,j} \times \boldsymbol{v}_{i,k}$ does not have the same direction with $\mathbf{n}$, then the triangular area is negative. We penalize the proportion of change in triangular area:

$$\mathcal{L}_{\text{fold}} = \frac{1}{T} \sum_{t=1}^{T} |\Delta_t - \Delta_t^0|, \Delta_t = \begin{cases} \Delta_t, & \text{if } \Delta_t \leq 0 \\ \Delta_t^0, & \text{otherwise} \end{cases}, \quad (10)$$

where $\Delta_t$ and $\Delta_t^0$ denote the current area and the original area of the triangle $t$, respectively, and a fold loss of zero indicates no folds or no self-intersections.

Overall, the regularization term consists of three distortion losses and a fold loss:

$$\mathcal{L}_{\text{reg}} = \lambda_{\text{areal}} \mathcal{L}_{\text{areal}} + \lambda_{\text{angle}} \mathcal{L}_{\text{angle}} + \lambda_{\text{dist}} \mathcal{L}_{\text{dist}} + \lambda_{\text{fold}} \mathcal{L}_{\text{fold}}, \quad (11)$$

where $\lambda_{\text{areal}}$, $\lambda_{\text{angle}}$, $\lambda_{\text{dist}}$, and $\lambda_{\text{fold}}$ are hyperparameters to balance weights of losses. $\lambda_{\text{fold}}$ takes much higher weight than others, due to the highest priority for preserving topology.

Additionally, anatomical segmentation or parcellation showed improved registration accuracy (Balakrishnan et al., 2019; Zhao et al., 2021a). Here, we leverage the Dice score of anatomical parcellation between the moved and fixed spheres to measure the registration accuracy in parcellation:

$$\text{Dice}(M^p \circ \boldsymbol{\Phi}, F^p) = \frac{2 \cdot |M^p \circ \boldsymbol{\Phi} \cap F^p|}{|M^p \circ \boldsymbol{\Phi}| + |F^p|}, \quad (12)$$

where $M^p_{\text{rigid}} \circ \Phi$ and $F^p$ denote the parcel $p$ on the moved and fixed spheres. The parcellation loss can be written as:

$$\mathcal{L}_{\text{parc}} = \frac{1}{P}\sum_{p=1}^{P}\left(1 - \text{Dice}(M^p \circ \Phi, F^p)\right), \tag{13}$$

where $P$ is the number of parcels, and the parcellations we used for model training are generated from FreeSurfer.

Taken together, for rigid registration, the loss function only uses the similarity term:

$$\mathcal{L}_{\text{rigid}} = \mathcal{L}_{\text{sim}}. \tag{14}$$

For non-rigid registration, we incorporate all aforementioned loss functions:

$$\mathcal{L}_{\text{non-rigid}} = \lambda_{\text{sim}}\mathcal{L}_{\text{sim}} + \mathcal{L}_{\text{reg}} + \lambda_{\text{parc}}\mathcal{L}_{\text{parc}}, \tag{15}$$

where $\lambda_{\text{sim}}$, $\lambda_{\text{reg}}$, and $\lambda_{\text{parc}}$ are weights to determine the relative importance between these terms.

## 2.6 Accelerated barycentric interpolation

The barycentric interpolation is widely used to interpolate values across triangular mesh surfaces, such as different resolutions of icospheres. However, repeated triangular face searching for the barycentric interpolation in the multi-resolution approach hampers the processing speed (Zhao et al., 2021a). To address this issue, we propose an accelerated barycentric interpolation by optimizing the procedure of triangular face searching. Specifically, we use the K-nearest neighbor (KNN) to locate the triangular faces containing $\kappa$ nearest vertices of the vertex to be interpolated $v_{\text{interp}}$ (Arefin et al., 2012). Here, we empirically set $\kappa = 8$ for each vertex, we account for up to 12 adjacent triangular faces for each of the nearest vertices. First, for each of adjacent triangular faces, we find the intersection point between the plane where the triangle located and the point to be interpolated $v_{\text{interp}} \in S^2$. Particularly, the intersection point is calculated by $v_{\text{intersect}} = \frac{\langle v_1, \mathbf{n} \rangle}{\langle v_{\text{interp}}, \mathbf{n} \rangle} v_{\text{interp}}$, where $v_1$ is a vertex of the triangle and $\mathbf{n}$ is the normal vector of the plane. Second, we examine if the $v_{\text{intersect}}$ located within the triangular face by calculating the $\omega$ for each vertex of a triangular face:

$$\omega_i = \langle v_{i,j} \times v_{i,\text{intersect}}, v_{i,j} \times v_{i,k} \rangle, \tag{16}$$

where $i, j, k \in \{1,2,3\}$, $i \neq j \neq k$ and $v$ denotes the vertex of the triangle. If and only if the $\omega$ for all three vertices are greater than 0, the $v_{\text{intersect}}$ locate within the triangular face. After determining which triangular face the intersection point locates within, we perform the barycentric interpolation using the three vertices of the face (Yeo et al., 2009). The accelerated barycentric interpolation is implemented using the PyTorch3D and thus could be performed on both CPU and GPU.

## 3. Experiments

To examine the registration performance in computational efficiency, accuracy, distortion, and test-retest reliability, we performed extensive experiments on 7 datasets consisting of over 10,000 scans.

### 3.1 Experimental setup
#### 3.1.1 Datasets

To enrich the diversity of MRI scanners, subjects' ages and ethnicities in the training set, we trained and validated the model in publicly available brain images of 904 subjects (labeled as "904" dataset in the current study) from the Consortium for Reliability and Reproducibility (CoRR)(Zuo et al., 2014) and the Southwest University Adult Lifespan Dataset (SALD) (Wei et al., 2018). The dataset was randomly split into 80% and 20% for training and validation. We further evaluated the registration performance of SUGAR on four unseen public datasets, the Washington University Alzheimer's Disease Research Center (ADRC) dataset, the Human Connectome Project Young Adult Unrelated 100 (HCP) dataset (Van Essen et al., 2013), the Midnight Scanning Club (MSC) dataset (Gordon et al., 2017), and a separate CoRR dataset collected at Hangzhou Normal University (CoRR-HNU). Of note, the CoRR-HNU is not a part of the CoRR dataset. Specifically, to assess the performance in registration accuracy and the amount of distortion, we used the ADRC dataset in which all images were manually annotated into 34 cortical areas by experts on each of the 39 subjects from the ADRC dataset. Moreover, we also assessed the performance of SUGAR on a subgroup of the HCP dataset including 100 subjects. Furthermore, to examine the test-retest reliability of our registration model, we used the MSC and CoRR-HNU datasets in which each subject underwent four or ten repetitive anatomical scans. Additionally, to intuitively demonstrate the application to the large-scale datasets, we applied our method to 9,000 subjects from the UK

BioBank dataset (Littlejohns et al., 2020; Miller et al., 2016). Of note, none of the aforementioned test datasets for performance evaluation was included in either the training or validation process.

***CoRR dataset.*** This dataset included data from 418 healthy subjects (228 females; 190 males; 6-62 years old). Images were collected on different 3T MRI scanners, consisting of Siemens TrioTim (Siemens Healthcare), GE Signa HDxt, and GE Discovery MR750 (GE Healthcare System) at voxel resolutions varying from 0.9 to 1.3 mm.

***SALD dataset.*** We used data from 486 subjects from the SALD dataset (304 females; 182 males; 19–80 years old). The T1w images were acquired on a Siemens TrioTim 3T MRI scanner using a magnetization-prepared rapid gradient echo (MPRAGE) sequence (repetition time (TR) = 1.9 s, echo time (TE) = 2.52 ms, flip angle = 90°, voxel size = $1\times1\times1$ mm$^3$).

***ADRC dataset.*** The dataset included 39 subjects acquired on a 1.5T Vision system (Siemens, Erlangen Germany). T1-weighted MPRAGE scans were obtained according to the following protocol: two sagittal acquisitions, TR = 9.7 ms, TE = 4 ms, flip angle = 10°, FOV = 224, and voxel size = $1\times1\times1.25$ mm$^3$. Two acquisitions were then averaged to increase the contrast-to-noise ratio. The cerebral cortex of each subject was manually parcellated into 34 cortical regions by a neuroanatomist and reviewed by two operators based on the depth of sulci and gyri.

***HCP dataset.*** We used data from 100 subjects from the HCP S900 release (54 females; 46 males; 22–80 years old). T1w images were acquired with a Siemens Connectome Skyra 3T MRI scanner using a submillimeter resolution MPRAGE sequence (TR = 2.4 s, TE = 2.14 ms, flip angle = 8°, FOV = 224, voxel size = $0.7\times0.7\times0.7$ mm$^3$).

***MSC dataset.*** We used data from 10 healthy adults from the MSC dataset (5 females; 5 males; 24–34 years old). Four T1w images were collected for each subject. The data was obtained on a Siemens TRIO 3T MRI scanner (TR = 2400 ms, TE = 3.74 ms, flip angle = 8°, voxel size = $0.8\times0.8\times0.8$ mm$^3$).

***CoRR-HNU dataset.*** We used data from 30 healthy adults from the CoRR-HNU dataset (15 females; 15 males; 20–30 years old). Each subject data point consisted of 10 scanning sessions across a one-month period. The T1w images were acquired on a GE Discovery MR750 3T MRI scanner equipped with an 8-channel head coil using a 3D "spoiled-gradient-echo" (SPGR) sequence (TR = 8.06 ms, TE = min full, flip angle = 8°, FOV = 250, voxel size = $1\times1\times1$ mm$^3$).

***UKB dataset.*** We used initial 9000 subjects from the UKB dataset (40-69 years old, 53% females). The T1w images were acquired on a 3T Siemens Skyra equipped with a 32-channel head coil using a 3D MPRAGE sequence (TR = 2000 ms, voxel size = 1×1×1 mm$^3$).

### 3.1.2 Data preprocessing

The ADRC, HCP, and UKB datasets were provided with officially preprocessed data using FreeSurfer. For the remaining datasets, we also used FreeSurfer to generate the required data for registration, including the spherical surfaces and geometric features, such as sulcal depth and curvature, in the native spaces with over 100,000 vertices per hemisphere. To fairly compare the performances among different non-rigid registration methods, we rigidly aligned all spheres to the atlas using SUGAR prior to non-rigid registration.

### 3.1.3 Baseline methods

The SUGAR model was compared against other five registration methods: FreeSurfer v7.1.0 release (Fischl et al., 1999a), Spherical Demons v1.5.1 (Yeo et al., 2009), MSM Pair (https://fsl.fmrib.ox.ac.uk/fsl/fslwiki/MSM) from FSL v6.0.5 (Robinson et al., 2014), MSM Strain (https://github.com/ecr05/MSM_HOCR) (Robinson et al., 2018), and S3Reg (https://github.com/zhaofenqiang/SphericalUNetPackage) (Zhao et al., 2021a). The FreeSurfer, Spherical Demons, MSM pair, and MSM Strain were implemented with their default settings. For fair comparison, we optimized the training parameters of S3Reg in the training and validation sets by performing a wide search on the combinations of smoothing terms for 200 training epochs, using a batch size of 1. Consequently, we selected the smoothing terms of 6, 12, 14 and 30 for four isosphere levels to train and generate an optimal model for subsequent comparisons.

### 3.1.4 Evaluation metrics

*Computational efficiency.* To evaluate the computational efficiency of SUGAR, we measured and compared the processing time of both rigid and non-rigid registration across the aforementioned baseline methods implemented on both CPU and GPU in an identical environment. For rigid registration, we compared the processing time between FreeSurfer and SUGAR. For non-rigid registration, we compared the processing time on CPU among FreeSurfer, Spherical Demons, MSM pair, MSM strain, S3Reg, and SUGAR. On GPU, we compared the processing time between S3Reg and SUGAR.

*Registration performance evaluation.* In terms of the registration accuracy, we first evaluated and compared parcellation accuracy of different methods to assess the similarity between the cortical anatomical parcellations derived from our approach and manually delineated annotations considered as the 'ground truth' (Yeo et al., 2009; Zhao et al., 2021a). We quantified the parcellation accuracy using the Dice score, where higher scores indicated greater similarity and better accuracy. Additionally, we assessed the alignment accuracy of geometric features, specifically the sulcal depth, by comparing it with the sulcal depth atlas from the FreeSurfer 'fsaverage' template surface. We employed the normalized cross-correlation (NCC) and mean absolute error (MAE) metrics to measure the similarity, with higher NCC values or lower MAE values indicating better alignment accuracy in sulcal depth.

Moreover, we evaluated the presence of self-intersections through calculating the Jacobian determinant for each triangular face. A negative Jacobian determinant indicates the existence of the self-intersections or folds. Furthermore, we evaluated the distortion of a registered surface from three aspects: areal distortion, shape distortion, and edge distortion (Robinson et al., 2018; Suliman et al., 2022b). Areal distortion was quantified by the absolute difference in areal values $(|\log_2 J|)$. Shape distortion quantifies the shape changes (i.e. shearing) before and after registration $(\log_2 R)$. The $J$ and $R$ were calculated as $J = \lambda_1 \cdot \lambda_2$ and $R = \lambda_1 / \lambda_2$, where $\lambda_1$ and $\lambda_2$ denote eigenvalues of the deformation matrix of each triangular face. The edge distortion measured the change in edge length after registration $(\log_2 L_2/L_1)$.

Additionally, we recognized the significance of test-retest reliability in neuroimaging (Xu et al., 2023). To evaluate the test-retest reliability, we examined the reproductivity of registration results across multiple scanning sessions from the same subject. Specifically, we calculated the vertex-wise intra-class correlation (ICC) of sulcal depth obtained from these repeated sessions. A higher ICC value represents higher test-retest reliability, indicating greater reproductivity in the registered surfaces across multiple scanning sessions.

### 3.1.5 Implementation and Training

Typically, registration to a single fixed atlas sphere is a conventional training strategy in previous learning-based methods (Cheng et al., 2020; Suliman et al., 2022a; Suliman et al., 2022b; Zhao et al., 2021a). However, the training strategy might restrict the learning capability of models without learning complex idiosyncrasies of the individual anatomies. To enhance the learning capability, we introduced a novel training strategy. In each epoch of the training process, a random sphere from the training set was selected as the registration target for each moving sphere, chosen from other subjects in the training set. Moreover, we augmented the resulting rigid registrations by adding three rotations along each rotation axis, with randomly generated rotation angles smaller than 0.01 radians. Taken together, these operations serve as a data augmentation strategy.

To simplify the comparisons among different registration methods as previous studies (Lyu et al., 2019; Suliman et al., 2022a), we focused on sulcal depth as the sole geometric feature for registration. It is important to note that our model is not constrained to this feature selection and can accommodate other types of features and even multimodal features in the registration process. Prior to inputting the cortical geometric features of the moving and fixed spheres into the S-GAT, we performed z-score normalization.

To reduce computational burden, we trained our model using a multi-resolution approach in a coarse-to-fine manner. The models were trained at 4 icosphere levels (i.e. $ico_3$, $ico_4$, $ico_5$, and $ico_6$). The upsampling and downsampling were performed through using the accelerated barycentric interpolation. The weights for four levels in loss functions were: $\lambda_{sim}$=[1, 1, 1, 1.5], $\lambda_{areal}$ = [1.5, 1.5, 1.5, 1.5], $\lambda_{angle}$=[1, 1, 1, 1], $\lambda_{dist}$=[2, 2, 2, 2], $\lambda_{fold}$=[30, 30, 35, 35], and $\lambda_{parc}$=[4, 4, 4, 4]. For the parcellation loss in the loss function, we used FreeSurfer to generate parcellation labels (i.e. 34 cortical areas) for each subject.

SUGAR was developed based on the PyTorch Geometric library (Fey and Lenssen, 2019) and optimized using the AdamW optimizer with a learning rate of $3 \times 10^{-4}$ for 200 training epochs. All of our experiments were executed on a workstation with an Intel Core i9 10980XE 3.00GHz × 36 CPU and a NVIDIA RTX3090 GPU with 24 GB RAM.

### 3.2. Comparison in computational efficiency across different registration methods

To compare computational efficiency of SUGAR with other registration methods, we measured processing time of each method for rigid and/or non-rigid registrations in the ADRC dataset (Table 1). For rigid registration, SUGAR took only 0.160 ± 0.009 seconds (mean ± std.) for each scan on CPU, approximately 340 times faster than FreeSurfer (54.315 ± 4.110 seconds), and its speed was nearly 9 times faster on GPU than on CPU. Similarly, for non-rigid registration, SUGAR significantly outperformed other registration methods in processing time on both CPU and GPU. On CPU, SUGAR ran approximately 18 times faster than the state-of-the-art conventional method, SD. On GPU, it was approximately 38 times faster than S3Reg. Taken together, SUGAR demonstrated a substantial improvement in computational efficiency, yielding an average total processing time of 0.235 seconds and 2.2 seconds for both rigid and non-rigid registration on GPU and CPU, respectively.

Table 1. Comparison of computational efficiency between different registration methods

|  |  | CPU Time (s) | GPU Time (s) |
|---|---|---|---|
| Rigid registration | FreeSurfer | 54.315 ± 4.110 *** | - |
|  | SUGAR | **0.160 ± 0.009** | 0.018 ± 0.013 |
| Non-rigid registration | FreeSurfer | 1885.999 ± 419.287*** | - |
|  | SD | 35.973 ± 8.356*** | - |
|  | MSM Pair | 753.927 ± 33.814*** | - |
|  | MSM Strain | 2129.364 ± 199.898*** | - |
|  | S3Reg | 12.540 ± 0.133*** | 8.309 ± 0.052*** |
|  | SUGAR | **2.058 ± 0.033** | **0.217 ± 0.022** |

Note: Two-tailed paired-sample *t* tests were conducted on processing time between SUGAR and other registration methods (***FDR-corrected $p < 0.001$). The numbers in bold indicate best performance in computational efficiency.

### 3.3. Registration performance evaluation

Given the trade-off between registration accuracy and distortion, we included parcellation accuracy and alignment accuracy of sulcal depth for evaluating the registration accuracy, while areal, shape, and edge distortions served for assessing distortions resulting from deformations. The performance in both accuracy and distortion were examined in two unseen datasets, namely the ADRC and HCP datasets. To investigate the reproductivity of different registration methods, we analyzed their performance in terms of test-retest reliability using the MSC and CoRR-HNU datasets. Additionally, to demonstrate the performance of SUGAR in large-scale datasets, we compared accuracy, distortion, and processing time of SUGAR with FreeSurfer in a 9000-subject subset of the UKB dataset.

#### 3.3.1. Registration accuracy

To evaluate the parcellation accuracy, we quantified similarity measured by overall Dice scores between manually annotated parcellation and automatic parcellation from different registration methods for FreeSurfer rigid and non-rigid registration, SD, MSM Pair, MSM Strain, S3Reg, and SUGAR rigid and non-rigid registration in the ADRC dataset (Table 2 and Table 3). For rigid registration, SUGAR showed comparable accuracy with FreeSurfer (Table 2). For non-rigid registration, SUGAR showed significantly higher accuracy than all conventional methods (Table 3, FDR-corrected $p$'s < 0.001) and higher but not statistically significant than S3Reg, demonstrating a good performance of both rigid and non-rigid registration in parcellation accuracy. To further evaluate parcellation accuracy in each of specific anatomical regions, we used the registration results from FreeSurfer as benchmarks and calculated the differences in Dice scores of each region between FreeSurfer and other methods (Figure 3). We observed that 8 cortical areas derived by SUGAR showed significantly larger Dice values than results from FreeSurfer (FDR-corrected $p$'s < 0.05) and none of the negative values showed statistically significant, which outperformed all other registration methods.

Table 2. Comparison of rigid registration accuracy between FreeSurfer and SUGAR

|  |  | Dice | NCC | MAE |
|---|---|---|---|---|
| ADRC dataset | FreeSurfer | 0.776 ± 0.045 | 0.732 ± 0.073** | 0.314 ± 0.040** |
|  | SUGAR | **0.778 ± 0.052** | **0.738 ± 0.067** | **0.313 ± 0.039** |
| HCP dataset | FreeSurfer | - | 0.683 ± 0.073** | 0.346 ± 0.037** |
|  | SUGAR | - | **0.692 ± 0.062** | **0.344 ± 0.036** |

Note: NCC = normalized cross-correlation; MAE = mean absolute error. Two-tailed paired-sample $t$ tests were conducted on accuracy indices between FreeSurfer and SUGAR (**FDR-corrected $p$ < 0.01). The numbers in bold indicate best performance in computational efficiency.

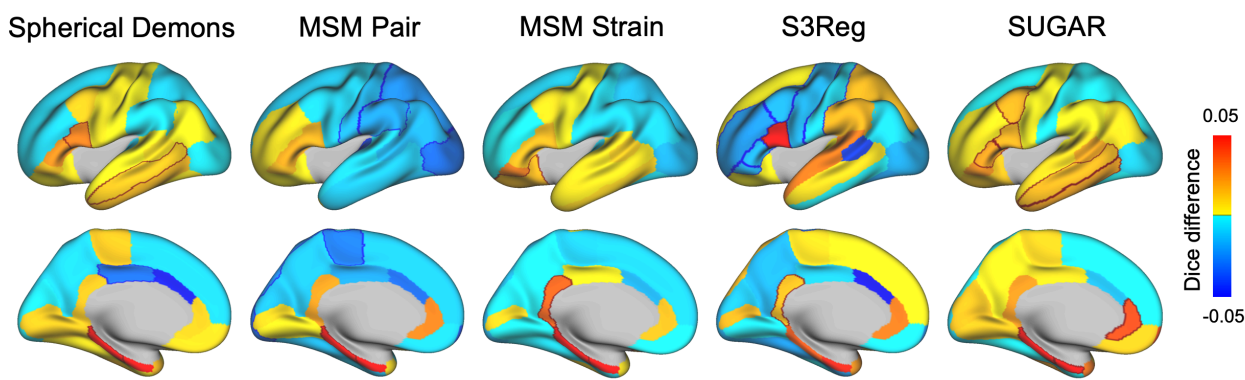

**Figure 3. Comparison of parcellation accuracy across different methods.** The Dice scores of anatomical parcellation between the manual annotations and automatically generated parcellation derived from different registration methods. For better illustration, we used the FreeSurfer as the benchmark method. Warm colors indicate parcels with higher Dice scores compared to FreeSurfer, and cool colors indicate parcels with lower Dice score compared to FreeSurfer. The dark red and dark blue lines highlight areas with significantly higher and lower Dice scores than FreeSurfer, respectively. Our model yields 84.85% of parcels with higher Dice scores and no areas significantly lower than FreeSurfer, outperforming other methods.

To further compare alignment accuracy of sulcal depth, we calculated NCC and MAE values between the sulcal depth atlas and sulcal depth derived from different registration methods. The ADRC dataset (Table 2 and Table 3) and another large unseen dataset, the HCP dataset (Table 4) were used to confirm our results. For rigid registration, SUGAR demonstrated significantly higher NCC and lower MAE compared to FreeSurfer in both the ADRC and the HCP datasets (FDR-corrected $p$'s < 0.001). For non-rigid registration, SUGAR consistently showed a significantly higher NCC and lower MAE than all conventional registration methods (FDR-corrected $p$'s < 0.01). SUGAR also showed higher NCC and lower MAE, albeit statistically non-significant, compared to S3Reg, except for the NCC in the HCP dataset (FDR-corrected $p$ < 0.05). These consistent findings highlight the improved alignment accuracy achieved by SUGAR. Importantly, our results also indicate the generalizability of our model, as it performed effectively on previously unseen datasets.

Table 3. Comparison of registration accuracy and spherical distortion between different registration methods (the ADRC dataset)

|  | Dice | NCC | MAE | Areal distortion | Shape distortion | Edge distortion |
|---|---|---|---|---|---|---|
| FreeSurfer | 0.827 ± 0.022*** | 0.874 ± 0.018*** | 0.213 ± 0.013*** | 0.290 ± 0.024*** | 0.486 ± 0.026*** | 0.200 ± 0.013*** |
| SD | 0.830 ± 0.022*** | 0.887 ± 0.017*** | 0.201 ± 0.013*** | 0.257 ± 0.019*** | *0.360 ± 0.016*** * | 0.163 ± 0.010*** |
| MSM Pair | 0.824 ± 0.022*** | *0.901 ± 0.013*** | *0.192 ± 0.011*** | 0.404 ± 0.050*** | 0.612 ± 0.060*** | 0.254 ± 0.026*** |
| MSM Strain | *0.832 ± 0.022*** | 0.889 ± 0.019*** | 0.199 ± 0.015*** | *0.173 ± 0.019*** | 0.423 ± 0.016*** | *0.161 ± 0.006*** |
| S3Reg | 0.825 ± 0.048 | 0.899 ± 0.031 | 0.196 ± 0.024 | 0.376 ± 0.021*** | 0.559 ± 0.034*** | 0.252 ± 0.013*** |
| SUGAR | **0.839 ± 0.021** | **0.904 ± 0.009** | **0.189 ± 0.009** | **0.166 ± 0.033** | **0.309 ± 0.035** | **0.121 ± 0.017** |

Note: NCC = normalized cross-correlation; MAE = mean absolute error. Two-tailed paired-sample *t* tests were conducted on the registration performance indices between SUGAR and other registration methods (**FDR-corrected $p < 0.01$, ***FDR-corrected $p < 0.001$). The numbers in bold indicate best performance and the numbers in italic indicate second-best performance in registration performance evaluation indices.

Table 4. Comparison of registration accuracy and spherical distortion between different registration methods (the HCP dataset)

|  | NCC | MAE | Areal distortion | Shape distortion | Edge distortion |
|---|---|---|---|---|---|
| FreeSurfer | *0.884 ± 0.015*** | 0.207 ± 0.013*** | 0.362 ± 0.024*** | 0.628 ± 0.033*** | 0.238 ± 0.011*** |
| SD | 0.873 ± 0.020*** | 0.214 ± 0.016*** | 0.285 ± 0.016*** | *0.391 ± 0.015*** | 0.174 ± 0.008*** |
| MSM Pair | 0.883 ± 0.016*** | 0.212 ± 0.013*** | 0.444 ± 0.041*** | 0.663 ± 0.054*** | 0.274 ± 0.020*** |
| MSM Strain | 0.864 ± 0.023*** | 0.223 ± 0.017*** | **0.185 ± 0.015*** | 0.444 ± 0.017*** | *0.160 ± 0.005*** |
| S3Reg | *0.884 ± 0.028* | 0.207 ± 0.022 | 0.374 ± 0.015*** | 0.559 ± 0.025*** | 0.252 ± 0.001*** |
| SUGAR | **0.891 ± 0.010** | **0.204 ± 0.009** | *0.191 ± 0.030* | **0.348 ± 0.039** | **0.136 ± 0.016** |

Note: NCC = normalized cross-correlation; MAE = mean absolute error. Two-tailed paired-sample *t* tests were conducted on the indices between SUGAR and other registration methods (*FDR-corrected $p < 0.05$, ***FDR-corrected $p < 0.001$). The numbers in bold indicate best performance and the numbers in italic indicate second-best performance in evaluation indices.

### 3.3.2. Registration distortions

To intuitively compare the distortion among different non-rigid registration methods, we first visualized spherical meshes after deformation from a randomly-selected subject in Figure 4. Overall, SUGAR yielded a smoother and less distorted spherical mesh than FreeSurfer, SD, MSM Pair, and S3Reg, and demonstrated comparable performance with the MSM Strain, which is specifically designed for minimizing distortions.

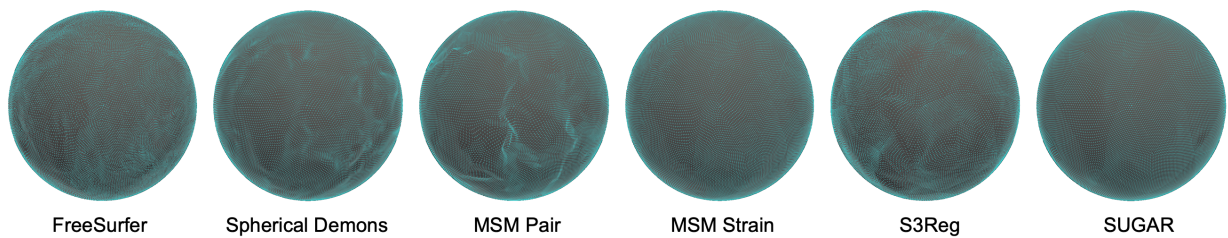

FreeSurfer    Spherical Demons    MSM Pair    MSM Strain    S3Reg    SUGAR

**Figure 4: Deformed spherical meshes with distortions from different methods.** The deformed spherical meshes from different methods of a representative subject from the ADRC dataset are illustrated. Obvious distortion could be found in FreeSurfer, SD, MSM Pair, and S3Reg. MSM Strain and SUGAR achieved the smoothest deformation, with distortion that is hard to be discerned.

To quantitatively assess the surface distortions, we evaluated areal, shape, and edge distortions induced by registration across different methods. The group-average distortion maps for each type were depicted for both the ADRC (Figure 5) and HCP (Figure 6) datasets. SUGAR consistently demonstrated low distortions across all three types while maintaining the highest registration accuracy. For the ADRC dataset, SUGAR exhibited significantly lower average areal, shape, and edge distortions compared to all other methods (FDR-corrected $p$'s < 0.01, Table 2). For the HCP dataset, SUGAR statistically outperformed FreeSurfer, SD, MSM Pair, and S3Reg in all types of distortions (FDR-corrected $p$'s < 0.001). It also outperformed MSM Strain in shape and edge distortions (FDR-corrected $p$'s < 0.001) and showed mildly lower average areal distortions (FDR-corrected $p$ < 0.05, Table 3). Our results indicate that SUGAR significantly reduced distortions while maintaining the outperforming registration accuracy when compared to other registration methods. Furthermore, it achieved comparable, if not better, distortion performance compared to MSM Strain, suggesting that SUGAR strikes a favorable balance between registration accuracy and distortions.

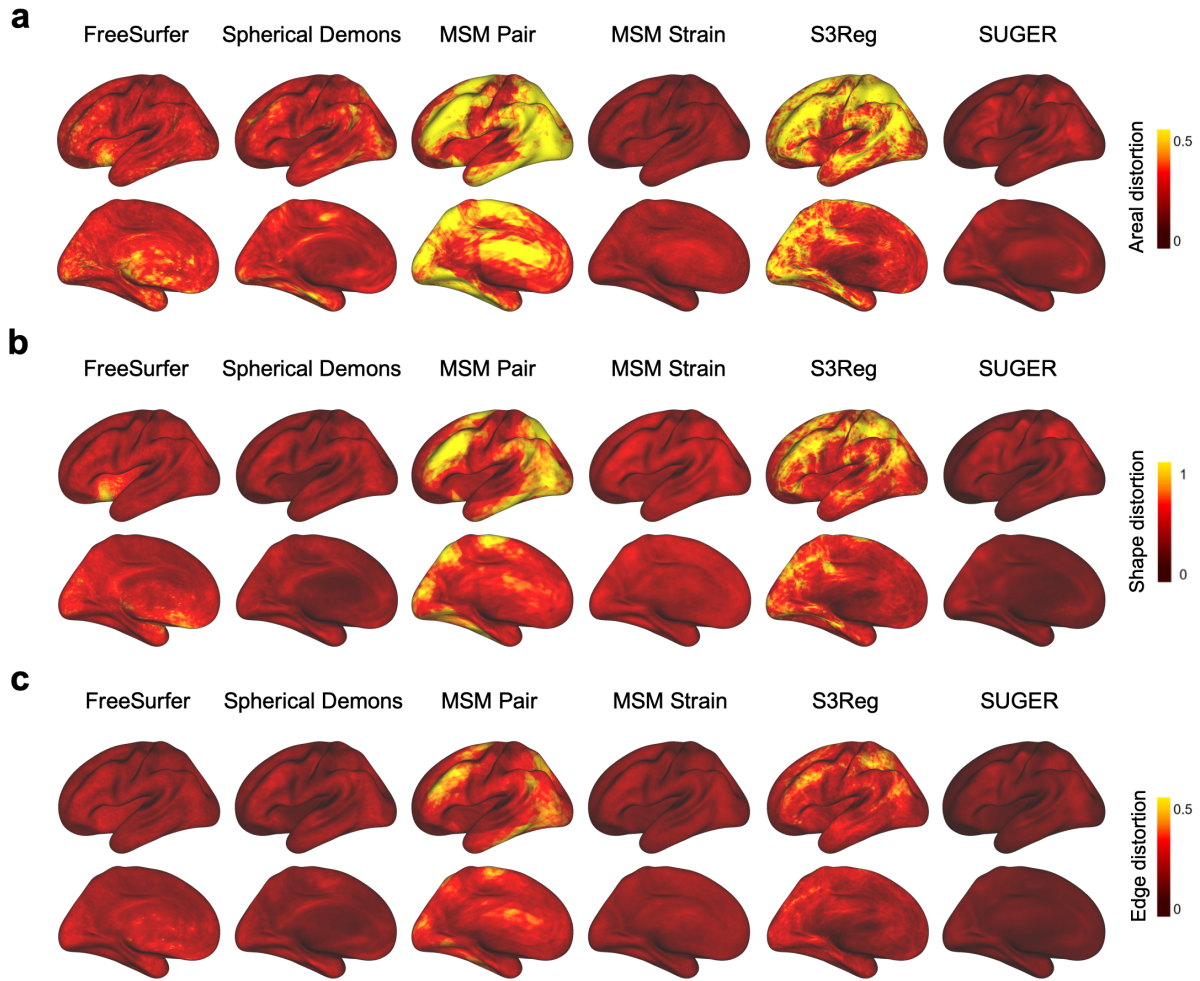

**Figure 5: The average distortion maps across participants on the ADRC dataset.** Group-average (a) areal, (b) shape, and (c) edge distortions from different methods are demonstrated. Darker colors indicate smaller distortions. SUGAR shows the least distortions in all types among all methods.

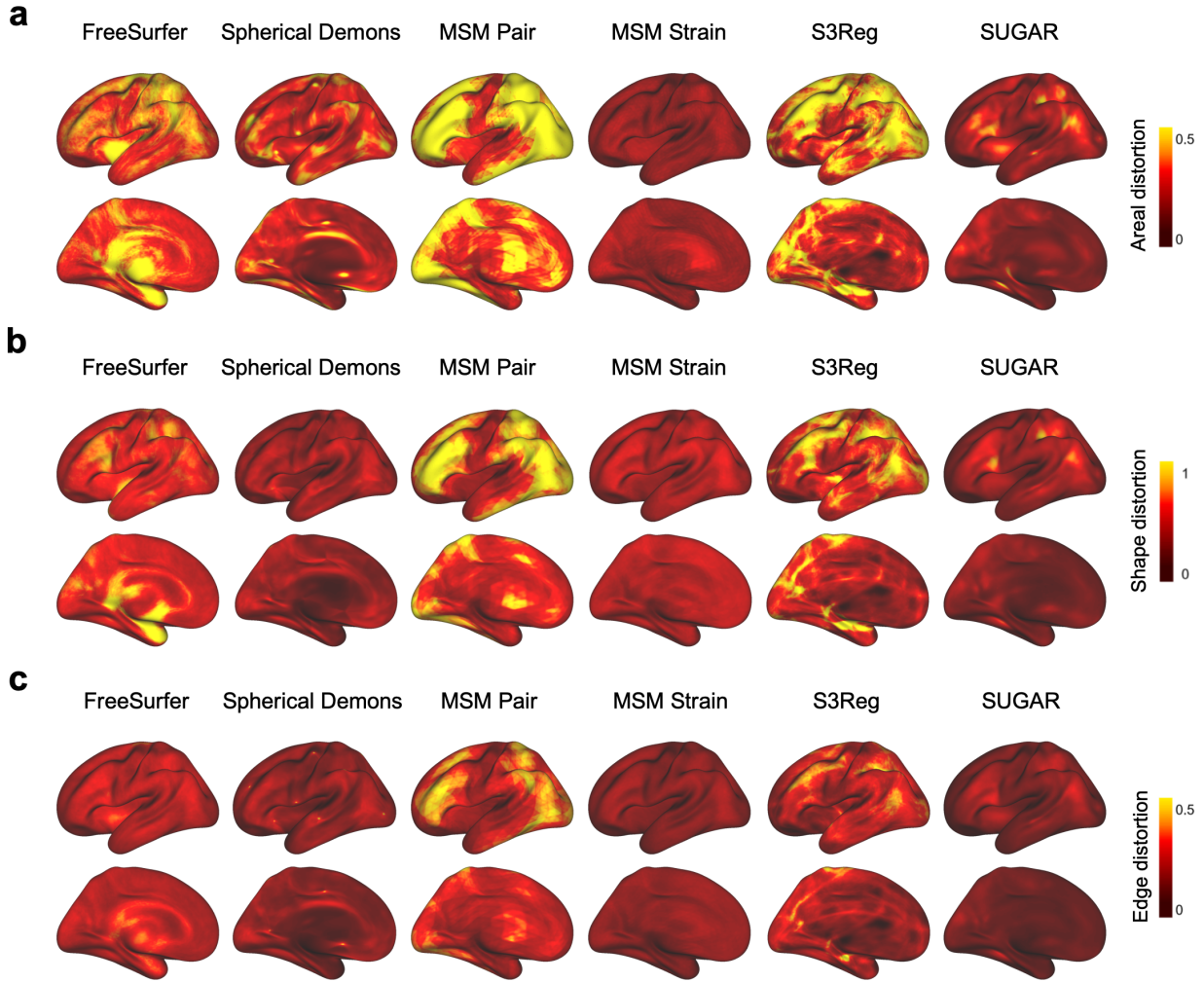

**Figure 6: The average distortion maps across participants on the HCP dataset.** MSM strain shows the least and the most even distributed areal distortions among all methods. SUGAR shows the least shape and edge distortions among all methods.

### 3.3.3. Test-retest reliability

In two unseen datasets with multiple anatomical scans for each subject (the MSC and CoRR-HNU dataset), we examined the test-retest reliability among different methods. The intra-class correlation maps and histogram distributions in sulcal depth were illustrated in Figure 7. We observed the highest test-retest reliability of FreeSurfer among all registration methods (the MSC dataset: mean (median) = 0.886 (0.947), 95% CI, 0.885–0.888; the CoRR-HNU dataset: 0.889 (0.939), 95% CI, 0.888–0.890). Despite lower reliability than FreeSurfer, SUGAR (the MSC dataset: 0.869 (0.932), 95% CI, 0.867–0.871; the CoRR-HNU dataset: 0.854 (0.903), 95% CI, 0.853–0.856) showed modestly improved registration compared to SD (the MSC dataset: 0.861

(0.919), 95% CI, 0.860–0.863; the CoRR-HNU dataset: 0.853 (0.904), 95% CI, 0.852–0.855), and outperformed other registration methods in both datasets (MSM Pair: the MSC dataset: 0.658 (0.697), 95% CI, 0.657–0.660; the CoRR-HNU dataset: 0.678 (0.701), 95% CI, 0.676–0.679; MSM Strain: the MSC dataset: 0.837 (0.891), 95% CI, 0.836–0.839; the CoRR-HNU dataset: 0.833 (0.871), 95% CI, 0.832–0.834; S3Reg: the MSC dataset: 0.788 (0.831), 95% CI, 0.786–0.789; the CoRR-HNU dataset: 0.743 (0.769), 95% CI, 0.742–0.745). These results suggested that SUGAR achieved comparable test-retest reliability with the widely used FreeSurfer, and significantly higher reliability than the learning-based method.

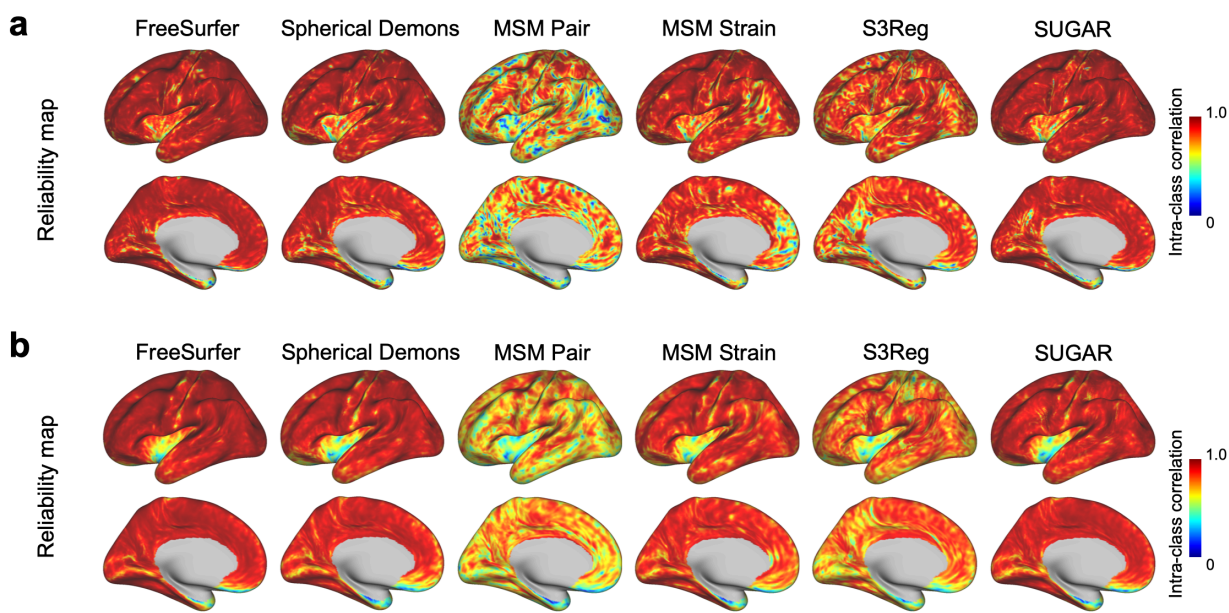

**Figure 7: Comparison of test-retest reliability in sulcal alignment.** (a) The test-retest reliability in sulcal alignment is estimated in the MSC dataset, in which each of 10 subjects underwent four T1w scans. The brain maps show the group-average vertex-wise test-retest reliability from different methods. Darker colors indicate higher reliability. FreeSurfer shows the highest reliability, and SUGAR follows and are significantly higher than other methods. (b) The test-retest reliability is estimated in the CoRR-HNU dataset, in which each of 30 subjects underwent ten T1w scans. The reliability maps show consistent results with the MSC dataset.

### 3.3.4. Performance in a representative large-scale dataset

We demonstrated the performance of SUGAR in a representative large-scale dataset i.e. a 9000-subject subset of the UKB dataset. Consistent with aforementioned findings, SUGAR outperformed FreeSurfer in both registration accuracy and distortions (Table 5). Notably, SUGAR could accomplish non-rigid registration for all 9,000 participants in 32.57 minutes on GPU, which is approximately 12,000 times faster than the time required by FreeSurfer (6615.30 hours on CPU). Of note, the processing time of FreeSurfer was obtained from the reports of the processed data released by the UKB. The high performance and computational efficiency of SUGAR show promise in large-scale neuroimaging studies.

## 4. Discussion

Our study introduces a continuous unsupervised learning-based method for cortical surface registration, yielding comparable if not better performance against state-of-the-art conventional and learning-based methods in terms of computational efficiency, registration accuracy, distortion, and test-retest reliability. These performance benefits are attributed to various aspects of our proposed framework, including the representational capability of the S-GAT, the design of the loss functions, the data augmentation strategy, and the positional encoding. These contributions will be discussed in detail in subsequent sections.

### 4.1 Framework

Cortical spherical meshes, which consist of vertices connected by edges, inherently possess a graph structure. GNNs thus effectively process and propagate spatial information through the vertices and edges of the mesh. By incorporating attention mechanisms, the S-GAT effectively and flexibly learns deformations on the spherical meshes, resulting in outperforming registration performance. Moreover, while much attention has been given to optimizing non-rigid registration due to its significance and computational complexity, the processing time of learning-based non-rigid registration has been considerably reduced to the sub-second level. This processing time is much faster compared to the rigid registration step in FreeSurfer, which is typically used as a precursor to non-rigid registration. Consequently, combining conventional rigid registration with learning-based non-rigid registration limits overall processing time. To address this issue, we integrated both rigid and non-rigid registration within a unified framework by leveraging the Euler angle representation. The unified framework and accelerated barycentric interpolation enabled us to achieve sub-second processing time for the complete registration task. Furthermore, our GNN-based framework possesses the capability to handle the alignment of multimodal features, such as functional connectivity, task activations, myelin, and other geometric features. However, the selection and weighting of these multimodal features are complex and vary depending on the specific aims of the study. Here, our aim was to propose a framework and perform a fair comparison with other conventional methods, thus we focused on the widely-used sulcal feature for alignment. Additionally, the S-GAT employed in our study is a generic GNN that can be theoretically extended to other time-consuming tasks on cortical surface meshes, such as the surface reconstruction (Ma et al., 2023; Ren et al., 2022b) and parcellation (Zhao et al., 2021b). This flexibility and versatility of the proposed network hold promise for establishing a learning-based surface pipeline with high performance and computational efficiency.

**4.2 Positional encoding**

Accurately establishing spatial correspondence is essential in cortical surface registration, as it relies heavily on spatial information. While the S-GAT model leverages the local relative positional information of each node, its performance may be hindered by the absence of global relative positional information, which is also important. We thus incorporated PE to provide additional global relative spatial information in our framework. To assess the impact of PE, we conducted experiments comparing the registration performance of the model with and without PE using the ADRC dataset. Our findings reveal that the model augmented with PE exhibited significantly higher alignment accuracy and reduced various types of distortion (Table 6, FDR-corrected *p's* < 0.01), emphasizing the importance of PE in the model.

Table 5. Comparison of computational efficiency, registration accuracy, and spherical distortion (the UKB dataset)

|  | NCC | MAE | Areal distortion | Shape distortion | Edge distortion |
|---|---|---|---|---|---|
| FreeSurfer | 0.863 (0.863–0.864) | 4.085 (4.083–4.087) | 0.304 (0.303–0.304) | 0.492 (0.491–0.492) | 0.202 (0.202–0.202) |
| SUGAR | **0.899 (0.899–0.899)** | **4.028 (4.026–4.030)** | **0.192 (0.191–0.192)** | **0.351 (0.350–0.352)** | **0.138 (0.138–0.138)** |

Note: NCC = normalized cross-correlation; MAE = mean absolute error. The numbers in parentheses indicate 95% confidence interval across participants. The numbers in bold indicate best performance in evaluation indices.

Table 6. Ablation study for positional encoding (PE).

|  | Dice | NCC | MAE | Areal distortion | Shape distortion | Edge distortion |
|---|---|---|---|---|---|---|
| Without PE | 0.835 ± 0.020** | 0.893 ± 0.015*** | 0.196 ± 0.012*** | 0.205 ± 0.030*** | 0.343 ± 0.034*** | 0.137 ± 0.016*** |
| Complete model | **0.839 ± 0.020** | **0.904 ± 0.009** | **0.189 ± 0.009** | **0.166 ± 0.033** | **0.309 ± 0.035** | **0.121 ± 0.017** |

Note: NCC = normalized cross-correlation; MAE = mean absolute error. Two-tailed paired-sample $t$ tests were conducted on the indices between complete SUGAR model and the model without PE (**FDR-corrected $p < 0.01$, ***FDR-corrected $p < 0.001$). The numbers in bold indicate best performance in evaluation indices.

## 4.3 Loss function

In our framework, we incorporated explicit similarity term and regularization term into the loss function, thereby enabling a flexible trade-off between registration accuracy and distortion. First, we set much higher weights for the fold loss (15-35 times than similarity term and other distortion losses), since preserving topology remains a high priority in the model. Through an ablation study on the ADRC dataset, we observed that self-intersections were shown in the model without the fold loss but not in the model with the loss, when ensuring the same level of registration accuracy. If we restrict no self-intersections in the model without the fold loss, the overall performance in accuracy significantly degraded after (Dice = 0.807 ± 0.290, NCC = 0.791 ± 0.033, MAE = 0.279 ± 0.021) compared to the model with the loss (Dice = 0.839 ± 0.021, NCC = 0.904 ± 0.009, MAE = 0.189 ± 0.009), indicating the fold loss is not only particularly critical for topology preserving but also for registration performance. Furthermore, we observed that each adapted loss significantly reduced distortion, with the areal loss primarily constraining areal distortion and the angular loss primarily constraining shape and edge distortions (Table 7, FDR-corrected $p$'s < 0.001), with the adjustment of registration accuracy to the similar level (mean Dice = 0.831~0.835, mean NCC = 0.899~0.904, mean MAE = 0.189~0.194). Overall, the inclusion of these losses played crucial roles in preserving topology and constraining distortion of deformed surfaces.

Table 7. Ablation study for areal, angular, and distance losses.

|  | Areal distortion | Shape distortion | Edge distortion |
|---|---|---|---|
| Without the three losses | 0.336 ± 0.023*** | 0.521 ± 0.027*** | 0.207 ± 0.012*** |
| Without areal loss | 0.200 ± 0.030*** | 0.320 ± 0.032*** | 0.127 ± 0.016*** |
| Without angular loss | 0.194 ± 0.032*** | 0.330 ± 0.033*** | 0.129 ± 0.016*** |
| Without distance loss | 0.173 ± 0.031*** | 0.314 ± 0.033*** | 0.123 ± 0.016*** |
| Complete model | **0.166 ± 0.033** | **0.309 ± 0.035** | **0.121 ± 0.017** |

Note: NCC = normalized cross-correlation; MAE = mean absolute error. Two-tailed paired-sample $t$ tests were conducted on the indices between complete SUGAR model and the model without a specific loss (***FDR-corrected $p$ < 0.001). The numbers in bold indicate best performance in evaluation indices.

## 4.4 Data augmentation strategy

Previous learning-based models in cortical surface registration commonly trained through aligning subjects' spheres to a fixed atlas. However, this training approach may limit the registration performance and robustness. We thus proposed the data augmentation strategy through randomized targets and rotations. To investigate whether this strategy improves registration performance and robustness, we conducted a comparative analysis of accuracy and distortions between models trained with and without the data augmentation strategy in the ADRC dataset. The results revealed significantly lower alignment accuracy of sulcal depth and larger distortions (FDR-corrected $p$'s < 0.001, Table 8) in the absence of the data augmentation strategy, highlighting its crucial role in enhancing model performance. Moreover, we also found the higher test-retest reliability of the model trained with the data augmentation strategy than that without the strategy in the MSC dataset (Table 8). Notably, the randomized atlas strategy not only benefited subject-to-atlas registration performance, but also extended the model's capability beyond subject-to-atlas registration tasks to subject-to-subject pairwise registration tasks (FDR-corrected $p$'s < 0.001, Table 9)

Table 8. Ablation study for the data augmentation strategy (DA).

| | Dice | NCC | MAE | Areal distortion | Shape distortion | Edge distortion | Test-retest reliability |
|---|---|---|---|---|---|---|---|
| Without DA | 0.834 ± 0.029 | 0.875 ± 0.033*** | 0.214 ± 0.021*** | 0.200 ± 0.026*** | 0.323 ± 0.035*** | 0.134 ± 0.015*** | 0.869 (0.867–0.870) |
| Complete model | **0.839 ± 0.020** | **0.904 ± 0.009** | **0.189 ± 0.009** | **0.166 ± 0.033** | **0.309 ± 0.035** | **0.121 ± 0.017** | **0.886 (0.885–0.888)** |

Note: NCC = normalized cross-correlation; MAE = mean absolute error. Two-tailed paired-sample $t$ tests were conducted on the indices between complete SUGAR model and the model without data augmentation (***FDR-corrected $p < 0.001$). The numbers in parentheses indicate 95% confidence interval across vertices. The numbers in bold indicate best performance in evaluation indices.

Table 9. Ablation study for the data augmentation strategy (DA) in subject-to-subject registration.

| | Dice | NCC | MAE | Areal distortion | Shape distortion | Edge distortion |
|---|---|---|---|---|---|---|
| Without DA | 0.784 (0.782–0.786) | 0.761 (0.759–0.763) | 0.284 (0.283–0.285) | 0.225 (0.224–0.227) | 0.357 (0.355–0.358) | 0.149 (0.148–0.149) |
| Complete model | **0.805 (0.804–0.807)** | **0.857 (0.857–0.858)** | **0.216 (0.216–0.217)** | **0.234 (0.232–0.236)** | **0.433 (0.431–0.435)** | **0.170 (0.169–0.171)** |

Note: NCC = normalized cross-correlation; MAE = mean absolute error. The numbers in parentheses indicate 95% confidence interval across participant pairs. The numbers in bold indicate best performance in evaluation indices.

## 4.5 Limitations and future directions

Although our proposed method achieved a highly computational efficient and comparable if not better registration performance against state-of-the-art methods, there are still several caveats and limitations that warrant attention. Firstly, the diffeomorphic registration is a one-to-one mapping without topological errors, which has been pursued by previous studies (Dalca et al., 2019; Yeo et al., 2009; Zhao et al., 2021a). However, recent studies have suggested that diffeomorphic assumption may restrict the registration performance due to the highly variable cortical topographies (Suliman et al., 2022a; Suliman et al., 2022b; Thual et al., 2022). In our study, we did not enforce diffeomorphisms but employed an additional fold loss to preserve topology. Our results demonstrated no self-intersections or folds (positive Jacobian determinant) following registration, in line with the outcomes of diffeomorphic methods. Given the advantageous properties of diffeomorphisms for specific research goals (Krebs et al., 2019; Lyu et al., 2019; Yeo et al., 2009), the development of a diffeomorphic variant of SUGAR may be worth exploring. Secondly, Although Euler angles representing deformation on spheres have no need for additional constraints to ensure the deformed vertices stay on the surface and are beneficial for the unified rigid and non-rigid registration framework, the representation has been subject to criticism due to its discontinuity in *SO(3)* and susceptibility to the gimbal lock (Zhou et al., 2019). However, given the small magnitude of deformations in cortical registration, the discontinuity concern becomes negligible (Zhou et al., 2020), and this representation proves to be efficient and accurate in characterizing the deformations observed in our results. Lastly, our study focused on evaluating registration performance on the cortical surfaces of healthy adults. However, it is equally important to optimize registration models for neonates and clinical populations with neuroanatomical abnormalities, including heavily lesioned and distorted brains, in both developmental and clinical research settings (Ren et al., 2022a; Ren et al., 2022b; Wang et al., 2023). The optimization of such models should be further investigated and validated in future studies.

## 5. Conclusion

In this study, we have presented the SUGAR framework, a comprehensive solution for both rigid and non-rigid cortical surface registration. The framework demonstrates exceptional performance in preserving the topology of deformations while achieving state-of-the-art results in terms of computational efficiency, registration accuracy, distortion control, and test-retest reliability. The accelerated framework with outstanding registration performance holds significant potential for facilitating large-scale neuroimaging studies. Furthermore, the S-GAT architecture, combined with the incorporation of fold and distortion losses, as well as the introduced data augmentation strategy, offer a generalizable approach that can be applied to various tasks involving spherical mesh representations.

**Declaration of Competing Interest**

H.L. is the chief scientist of Beijing Neural Galaxy Technology Co., Ltd., which is not a sponsor of this study. Other authors declare no conflict of interest regarding the publication of this work.

**Acknowledgements**

This work was supported by Changping Laboratory (2021B-01-01) and the China Postdoctoral Science Foundation (2022M720529, 2023M730175). We acknowledge the National Center for Protein Sciences at Peking University for assistance with MRI data processing tools. The MRI data used for model training and evaluation in the preparation of this manuscript were provided by (1) the Consortium for Reliability and Reproducibility (Consortium Founders: Xi-Nian Zuo and Michael P. Milham), whose funding and manpower were provided by The National Institute on Drug Abuse (NIDA) and the National Natural Science Foundation of China (NSFC), along with the Child Mind Institute, the Institute of Psychology, the Chinese Academy of Sciences and the Nathan Kline Institute; (2) the Southwest University Adult Lifespan Dataset (Investigators: Jiang Qiu, Gongrong Wu, Dongtao Wei, Wenjing Yang, Qunlin Chen and Kaixiang Zhuang), supported by the National Natural Science Foundation of China (31470981; 31571137; 31500885), National Outstanding young people plan, the Program for the Top Young Talents by Chongqing, the Fundamental Research Funds for the Central Universities (SWU1509383, SWU1509451, SWU1609177), the Natural Science Foundation of Chongqing (cstc2015jcyjA10106), the Fok Ying Tung Education Foundation (151023), the General Financial Grant from the China Postdoctoral Science Foundation (2015M572423, 2015M580767), Special Funds from the Chongqing Postdoctoral Science Foundation (Xm2015037, Xm2016044), Key research for Humanities and Social Sciences of Ministry of Education (14JJD880009); (3) the Human Connectome Project, WU-Minn Consortium (Principal Investigators: David Van Essen and Kamil

Ugurbil; 1U54MH091657), funded by the 16 NIH Institutes and Centers that support the NIH Blueprint for Neuroscience Research; and by the Mc-Donnell Center for Systems Neuroscience at Washington University; (4) This MSC data was obtained from the OpenNeuro database. Its accession number is ds000224; (5) This research has been conducted using data from UK Biobank, a major biomedical database, with the application number of 99038 (www.ukbiobank.ac.uk).